\pdfoutput=1
\documentclass{appolb}

\usepackage{amsmath}
\usepackage{amsfonts}
\usepackage{graphicx}
\usepackage{slashed}
\usepackage{subcaption}

\usepackage{geometry}
\usepackage{pdflscape}

\usepackage{hyperref}

\DeclareMathOperator{\Order}{\mathcal{O}}

\newcommand{\Ca}{C_a}
\newcommand{\Cf}{C_f}
\newcommand{\D}{\mathrm{d}}
\newcommand{\Li}{\mathrm{Li_2}}
\newcommand{\Pgg}{P_{gg}}
\newcommand{\Pqq}{P_{qq}}
\newcommand{\Tf}{T_f}

\newcommand{\abs}[1]{\lvert #1 \rvert}
\newcommand{\as}{\alpha_\mathrm{s}}
\newcommand{\eps}{\epsilon}
\newcommand{\mr}{\mu_{r}}

\newcommand{\pqq}{p_{qq}}

\begin{document}
  \eqsec  
  
  \title{
  \begin{flushright}
    DESY 14-087 \\
    IFJPAN-IV-2014-10 \\
    LPN 14-077
  \end{flushright}
  Calculation of QCD NLO Splitting Functions in the light-cone gauge: a new regularization prescription 
  \thanks{Presented by O.Gituliar at XX Cracow Epiphany Conference, Cracow (Poland), 8--10 Jan 2014.}
  }
  
\author{ O.\ Gituliar$^{a,b}$, S.\ Jadach$^a$, A.~Kusina$^c$, M.\ Skrzypek$^a$
\address{$^a$ Institute of Nuclear Physics, Polish Academy of Sciences,\\
              ul.\ Radzikowskiego 152, 31-342 Cracow, Poland}
\address{$^b$ DESY, Platanenallee 6, D-15738 Zeuthen, Germany}
\address{$^c$ Southern Methodist University, Dallas, TX 75275, USA}
}

  \maketitle

\begin{abstract}
We report on the progress in calculating NLO DGLAP splitting functions for $x<1$
using the New Principal Value prescription, which is a modification of the
standard
Principal Value approach proposed by Curci, Furmanski and Petronzio in 1980.
The new prescription reproduces the standard results on the inclusive
(integrated)
level, but simplifies individual contributions and restricts the cancellations
between real and virtual diagrams which makes it useful for Monte Carlo
simulations.
\end{abstract}


\section{Introduction}

In QCD
in addition to the ultra-violet (UV) singularities also mass singularities
appear.
Mass singularities can be further divided into infra-red (IR) and collinear
ones.
By analogy with the UV divergences, which rule the evolution of the strong
coupling constant $\as$,
the collinear divergences rule the evolution of
parton distribution functions, $f_i$.
This evolution is governed by the DGLAP equations
\cite{AP77,Gri72,Dok77} which read:
\begin{equation} \label{eq:dglap}
  \frac{\partial}{\partial \ln\mu^2} f_i\left(x,\mu^2\right)
  =
  \sum_{j=g,q,\bar{q}}
  \int_x^1 \frac{\D z}{z}
P_{ij}\left(z,\as(\mu^2)\right)f_j\Bigl(\frac{x}{z},\mu^2\Bigr)
  \text{,}
\end{equation}
where $\mu$ is the factorization scale, $i,j=g,q,\bar{q}$ goes over all types of
partons,
and $P_{ij}\left(x,\as\right)$ are the DGLAP splitting functions (evolution
kernels).

As opposed to parton distributions, the evolution kernels can be calculated
perturbatively
and we can write their expansion in powers of $\as$ as
\begin{equation}
  P_{ij}\left(x,\as(\mu^2)\right)
  =
  \frac{\as}{2\pi} P_{ij}^{(0)}\left(x\right)
  +
  \Bigl(\frac{\as}{2\pi}\Bigr)^2 P_{ij}^{(1)}\left(x\right)
    + \Order(\as^3)
  \text{.}
\end{equation}
The lowest $\Order(\as)$ splitting functions $P_{ij}^{(0)}(x)$ are known for
about forty years, e.g.~see \cite{AP77,CFP80}.
Later, the $\Order(\as^2)$ splitting functions were calculated with the help of
two independent methods: 1) the operator product expansion
(OPE)~\cite{FRS77,FRS79} and 2) the method of Curci, Furmanski, and Petronzio
(CFP) \cite{CFP80,FP80} which is based on factorization properties of mass
singularities in the light-cone gauge \cite{EGMPR79}.
The latest, state-of-the-art results for the $\Order(\as^3)$ space-like
splitting functions were obtained with OPE approach in \cite{MVV04a,MVV04b},
and approximate results for the $\Order(\as^3)$ time-like splitting functions
are available in \cite{AMV12,MMV06}.

Such impressive results obtained with the OPE method are caused by its technical
simplicity compared to the CFP approach.
The latter method operates in the physical momentum space and uses light-cone
gauge which introduces additional complications due to the axial-type
denominator $1/(ln)$, where $n$ is a light-cone reference vector.
Despite of this complication, the CFP method provides a more clear physical
picture of the collinear limit, i.e. generalized ladder expansion
\cite{EGMPR79}.

This fact is used to build the first fully next-to-leading-order (i.e.
$\Order(\as^2)$) parton shower Monte Carlo
generator \cite{GJKPSS11,JKPSS13,JSKS09} (with NLO corrections included in both
the hard matrix element and the shower itself).
Construction of such a NLO shower, however, requires the knowledge about the
$\Order(\as^2)$ splitting functions at the exclusive, unintegrated, level.
It means that, in addition to the known dependence on the longitudinal momentum
fraction $x$, the dependence on the transverse momentum variable should be also
known.
Because of this, the use of the OPE method
becomes more complicated in comparison to the CFP approach.

Moreover, the $\Order(\as^2)$ splitting functions calculated in the standard CFP
approach,
that are available in the literature~\cite{CFP80,EV96,Hei98}, are also not well
suited
for the purpose of Monte Carlo simulations of parton cascades. First, the
literature provides mostly the
inclusive splitting functions (after integration over the final state momenta).
Second, in the classic
CFP approach there are cancellations between real and virtual contributions 
that are realized through
cancellation of dimensional $\epsilon$ poles. This is incompatible with the
Monte Carlo parton
shower algorithms, and motivates modification of the CFP
prescription~\cite{GJKS14} and recalculation
of the $\Order(\as^2)$ evolution kernels.%
    \footnote{There is also an interest in recalculating the evolution kernels
for the purpose
    of improving the convergence of expansion of parton
distributions~\cite{OMR13,Oliveira13}.}
    
We already started to calculate the NLO DGLAP kernels in the modified scheme.
For the first time the new approach was used to calculate the real contributions
to
the non-singlet splitting function \cite{JKSS11}. After that, we proceed
with one-loop
virtual contributions \cite{Git14} and would like to report on these results in
this paper.

\section{The New Principal Value Prescription}

Contributions to the NLO splitting functions for $x<1$ come from two types of
topologies, further referred to as {\em real} -- with two final-state momenta;
and {\em virtual} -- with one virtual and one final-state momentum.
At the inclusive level these two cases can be added together since all, real and
virtual, momenta are integrated out and only $x$-dependence is left.
With more detailed analysis of the standard calculation procedure \cite{Hei98}
one can find that $1/\eps^3$ terms cancel between these two contributions,
while, as expected, terms $1/\eps^2$ and lower survive.
Existence of $1/\eps^3$ poles is a serious problem for Monte-Carlo application
of splitting functions, since at the exclusive level the real and virtual
contributions are defined in different phase spaces, two- and one-particle ones
respectively.
Therefore, the contributions can not be easily added in order to cancel
higher-order poles in $\eps$ and to make splitting functions finite in
$\eps \to 0$ limit at the exclusive level.
Fortunately, this issue can be resolved with a modification to the standard PV
regularization prescription for the light-cone singularities, which we proposed
in \cite{GJKS14,Git14}.

In the standard approach the PV regularization is applied only to the axial
denominator
in the gluon propagator at the level of the Feynman rules:
\begin{equation}
g^{\mu\nu} - \frac{l^{\mu}n^{\nu}+l^{\nu}n^{\mu}}{ln}
\rightarrow
g^{\mu\nu} -
\left(l^{\mu}n^{\nu}+l^{\nu}n^{\mu}\right)\frac{ln}{(ln)^2+\delta^2(pl)},
\end{equation}
where $\delta$ is an infinitesimal regulator and $p$ is an external reference
momentum.
However, there are also singularities in the $l_+=(ln)/(pn)$ variable
originating from the
phase space parametrization or the Feynman part of gluon propagator and in the
standard
approach these are regularized 
in a dimensional manner.
In the {\em New Principal Value} (NPV) prescription~\cite{GJKS14}
the PV regularization is used to regularize {\em all} the singularities
in the $l^+$ variable, regardless of their origin.
For instance, we do the following replacement:
\begin{equation}
d^ml \,l_+^{\epsilon-1}
\rightarrow
d^ml \,\frac{l_+}{l_+^2+\delta^2} \left(1 + \epsilon \ln l_+ + \epsilon^2
\frac{1}{2} \ln^2l_+ + \dots \right)
\end{equation}
in the entire integrand.

When the new prescription is used, there is one technical point
we would like to underline. As opposed to the standard techniques of calculating
loop integrals, where we introduce e.g.\ the Feynman parameters and integrate
over
them at the very end of the calculation, in the NPV prescription we perform
the integration over the ``plus" component ($l_+$) as the last one.

In the NPV prescription singularities appearing as double $\epsilon$ poles
(before integration over scale, or in the case of virtual graphs before
integration
over final state momentum) are changed to double logarithms in the regulator
$\delta$ or combination of single logarithms and single poles.%
    \footnote{We use the following symbols for divergent integrals with the 
geometrical
    PV regularization:
    \[
    I_0 = \int_0^1 dx \frac{x}{x^2+\delta^2} \simeq -\ln\delta,
    \qquad
    I_1 = \int_0^1 dx \frac{x}{x^2+\delta^2}\ln x 
       \simeq -\frac{1}{2}\ln^2\delta-\frac{\pi^2}{24}.
    \]
}
For example, if we consider a three-point Feynman integral
(kinematical set-up: $p^2=(p-q)^2=0,\; q^2< 0$)
in the standard CFP approach, we have:
\begin{equation}
  J_3^\mathrm{F}
  =
  \int \frac{\D^m l}{(2\pi)^m}
    \frac{1}{l^2 (q-l)^2 (p-l)^2}
  =
  \frac{i}{(4\pi)^2 \abs{q^2}}
  \left(
    \frac{4\pi}{\abs{q^2}}
  \right)^{-\epsilon}
  \Gamma(1-\epsilon)
  \left(
    -\frac{1}{\eps^2} + \frac{\pi^2}{6}
  \right),
\label{eq:pv-f}
\end{equation}
and in the NPV prescription it is given by:
\begin{equation}
\begin{split} \label{eq:our-f}
  J_3^\mathrm{F}
  = &
  \frac{i}{(4\pi)^2 \abs{q^2}}
  \left(
    \frac{4\pi}{\abs{q^2}}
  \right)^{-\epsilon}
  \Gamma(1-\epsilon)
  \bigg(
   - \frac{2 I_0 + \ln(1-x)}{\eps}
    - 4 I_1 + 2 I_0 \ln(1-x) + \frac{\ln^2(1-x)}{2}
  \bigg).
\end{split}
\end{equation}
In this example we can explicitly see how the $\epsilon$ pole is
replaced by the geometrical regulator which can be actually implemented
in a Monte Carlo program.
It is also important that in most cases the singularities regularized
by $I_0$ and $I_1$ cancel separately between the real and virtual contributions.

\section{NLO Splitting Functions}

\begin{figure}[ht]
  \centering
  \subcaptionbox*{\centerline{(c):
$\Cf^2-\frac{1}{2}\Cf\Ca$}}{\includegraphics[scale=1.15]{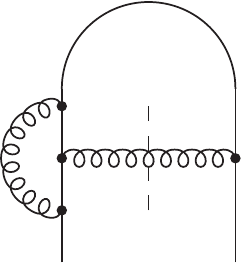}}
  \hspace{2cm}
  \subcaptionbox*{\centerline{(d-qq):
$\frac{1}{2}\Cf\Ca$}}{\includegraphics[scale=1.15]{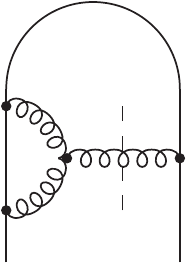}}
  \hspace{2cm}
  \subcaptionbox*{\centerline{(d-gg):
$\Ca^2$}\vspace{1cm}}{\includegraphics[scale=1.15]{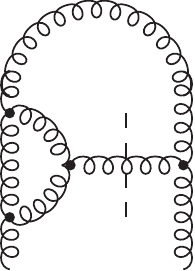}}
  \hspace{2cm}
  \subcaptionbox*{\centerline{\hspace{6mm}(e): $\Cf^2$}
}{\includegraphics[scale=1.15]{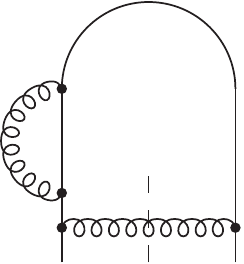}}
  \hspace{2cm}
  \subcaptionbox*{\centerline{(f):
$\Cf\Ca$}}{\includegraphics[scale=1.15]{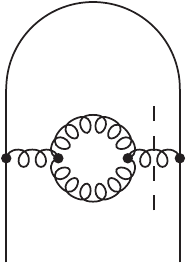}}
  \hspace{2cm}
  \subcaptionbox*{\centerline{(g):
$\Cf\Tf$}}{\includegraphics[scale=1.15]{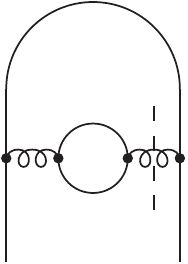}}
  \vspace{5mm}
  \caption{Virtual graphs contributing to the NLO
splitting function $P_{qq}$ and one contribution to the $P_{gg}$ kernel.}
  \label{fig:1}
\end{figure}

In this section, we present a few selected results obtained in the NPV
scheme. Namely, we show
all contributions to the inclusive splitting function $P_{qq}$ and
selected contributions to the $P_{gg}$ one.
 This
is enough to demonstrate explicitly that the NPV prescription
leads to the same inclusive splitting functions as the standard PV scheme,
but at the same time can be used at the exclusive level for Monte Carlo
simulations in four dimensions. All the contributing graphs are shown in Fig.\
\ref{fig:1}.

{\em Inclusive splitting functions} are presented in
Tables~\ref{tb:1}--\ref{tb:2}, where in the `virt' column we provide
contributions with one cut line, in the `real' column -- with two cut lines,
and in `sum' column -- their sum.
This allows for comparison of the results in NPV prescription with
already
known results in the PV prescription \cite{CFP80,EV96,Hei98} and demonstrates
the correctness of the proposed approach.
The results in the `virt' columns can be calculated with the help of the {\tt
Axiloop}
package \cite{Axiloop}
and are available in \cite{Git14,GJKS14}, the real contributions were
calculated in
\cite{JKSS11}.
The column `sum' equals to the sum of the corresponding `virt' and `real'
columns and is in full agreement with the known results, e.g. \cite{EV96,Hei98}.

\begin{table}
  \centering
  \begin{tabular}{|l||c|c|c||c|c|c||c|c|c|}
    \hline
    & \multicolumn{3}{|c||}{(d-qq)}
    & \multicolumn{3}{|c||}{(d-gg)}
    & \multicolumn{3}{|c|}{(c)}
    \\ \hline
                                      & virt   & real   & sum    & virt   & real
  & sum & virt   & real   & sum
    \\ \hline \hline
    \multicolumn{10}{c}{\em Double Poles}
    \\ \hline
    $\pqq$                            & $-6$   & $0$    & $-6$   & $-22/3$  &
$0$     & $-22/3$  & $-6$   & $0$    & $-6$
    \\ 
    $p_{qq} \; \ln{x}$                & $4$    & $0$    & $4$    & $4$      &
$0$     & $4$      & $4$    & $0$    & $4$
    \\ 
    $p_{qq} \; \ln(1-x)$              & $8$    & $0$    & $8$    & $4$      &
$0$     & $4$      & $0$    & $0$    & $0$
    \\
    $p_{qq} \; I_0$                   & $16$   & $0$    & $16$   & $12$     &
$0$     & $12$     & $8$    & $0$    & $8$
    \\ \hline \hline
\multicolumn{10}{c}{\em Single Poles}
    \\ \hline
    $p_{qq}$                          & $-7$   & $-4$   & $-11$  & $-134/9$ &
$-4$    & $-170/9$ & $-7$   & $0$    & $-7$
    \\
    $p_{qq} \; \ln{x}$                & $ 0$   & $-3/2$ & $-3/2$ & $0$      &
$-11/3$ & $-11/3$  & $0$    & $-3/2$ & $-3/2$
    \\
    $p_{qq} \; \ln(1-x)$              & $-3$   & $8$    & $5$    & $-22/3$  &
$8$     & $2/3$    & $-3$   & $0$    & $-3$
    \\
    $p_{qq} \; \ln^2{x}$              & $2$    & $-1$   & $1$    & $4$      &
$-2$    & $2$      & $2$    & $-1$   & $1$
    \\
    $p_{qq} \; \ln{x} \ln(1-x)$       & $2$    & $4$    & $6$    & $4$      &
$4$     & $8$      & $2$    & $0$    & $2$
    \\
    $p_{qq} \; \ln^2(1-x)$            & $4$    & $-2$   & $2$    & $4$    & $-2$
     & $2$      & $0$    & $0$    & $0$
    \\
    $p_{qq} \;\: \Li(1)$              & $8$    & $-2$   & $6$    & $12$   & $-2$
     & $10$     & $4$    & $0$    & $4$
    \\
    $p_{qq} \;\: \Li(1-x)$            & $-2$   & $2$    & $0$    & $0$    & $0$ 
     & $0$      & $2$    & $-2$   & $0$
    \\
    $1/x$                             & $0$    & $0$    & $0$    & $0$    &
$-22/3$   & $-22/3$  & $0$    & $0$    & $0$
    \\
    $1$                               & $-3$   & $2$    & $-1$   & $0$    &
$25/3$    & $25/3$   & $-3$   & $-8$   & $-11$
    \\
    $x$                               & $2$    & $-1$   & $1$    & $-1/3$ &
$-24/3$   & $-25/3$  & $4$    & $7$    & $11$
    \\
    $x^2$                             & $0$    & $0$    & $0$    & $0$    &
$22/3$    & $22/3$   & $0$    & $0$    & $0$
    \\
    $1/x \ln{x}$                      & $0$    & $0$    & $0$    & $0$    &
$-11/3$   & $-11/3$  & $0$    & $0$    & $0$
    \\
    $\ln{x}$                          & $2$    & $1/2$  & $5/2$  & $0$    &
$-23/6$   & $-23/6$  & $2$    & $-7/2$ & $-3/2$
    \\
    $x \ln{x}$                        & $-2$   & $1/2$  & $-3/2$ & $0$   &
$-23/6$    & $-23/6$  & $-2$   & $-7/2$ & $-11/2$
    \\
    $x^2 \ln{x}$                      & $0$    & $0$    & $0$    & $0$    &
$-11/3$   & $-11/3$  & $0$    & $0$    & $0$
    \\
    $(1-x) \; \ln(1-x)$               & $4$    & $0$    & $4$    & $0$    & $0$ 
     & $0$      & $0$    & $0$    & $0$
    \\ \hline
\multicolumn{10}{|c|}{\em Single Spurious Poles}
    \\ \hline
    $p_{qq} \; I_0$                   & $0$    & $8$    & $8$    & $0$    & $8$ 
     & $8$      & $0$    & $0$    & $0$
    \\
    $p_{qq} \; I_0 \ln{x}$            & $4$    & $4$    & $8$    & $8$    & $4$ 
     & $12$     & $4$    & $0$    & $4$
    \\
    $p_{qq} \; I_0 \ln(1-x)$          & $12$   & $-4$   & $8$    & $16$   & $-4$
     & $12$     & $4$    & $0$    & $4$
    \\
    $p_{qq} \; I_1$                   & $-12$  & $4$    & $-8$   & $-16$  & $4$ 
     & $-12$    & $-4$   & $0$    & $-4$
    \\
    $(1-x) \; I_0$                    & $8$    & $0$    & $8$    & $0$    & $0$ 
     & $0$      & $4$    & $0$    & $4$
    \\ \hline
  \end{tabular}
  \caption{Contributions to the inclusive splitting functions $P_{qq}$ and
$P_{gg}$ from the graphs (d-qq), (d-gg), and (c) in
the NPV prescription.}
  \label{tb:1}
\end{table}

\begin{table}
  \centering
  \begin{tabular}{|l||c||c|c|c||c|c|c|}
    \hline
    & (e)
    & \multicolumn{3}{|c||}{(f)}
    & \multicolumn{3}{|c|}{(g)}
    \\ \hline
                                      & virt & virt    & real    & sum     &
virt   & real    & sum
    \\ \hline \hline
    \multicolumn{8}{c}{\em Double Poles}
    \\ \hline
    $\pqq$                            & $6$  & $44/3$  & $-22/3$ & $22/3$  &
$-8/3$ & $4/3$   & $-4/3$
    \\ 
    $p_{qq} \; \ln{x}$                & $-8$ & $0$     & $0$     & $0$     & $0$
   & $0$     & $0$
    \\ 
    $p_{qq} \; \ln(1-x)$              & $0$  & $-16$    & $8$    & $-8$    & $0$
   & $0$     & $0$
    \\
    $p_{qq} \; I_0$                   & $-8$ & $-16$   & $8$     & $-8$    & $0$
   & $0$     & $0$
    \\ \hline \hline
\multicolumn{8}{c}{\em Single Poles}
    \\ \hline
    $p_{qq}$                          & $7$  & $0$     & $103/9$ & $103/9$ & $0$
   & $-10/9$ & $-10/9$
    \\
    $p_{qq} \; \ln{x}$                & $0$  & $0$     & $11/3$  & $11/3$  & $0$
   & $-2/3$  & $-2/3$
    \\
    $p_{qq} \; \ln(1-x)$              & $3$  & $22/3$  & $-34/3$ & $-4$    &
$-4/3$ & $4/3$   & $0$
    \\
    $p_{qq} \; \ln^2{x}$              & $-2$ & $0$     & $0$     & $0$     & $0$
   & $0$     & $0$
    \\
    $p_{qq} \; \ln{x} \ln(1-x)$       & $-4$ & $0$     & $-4$    & $-4$    & $0$
   & $0$     & $0$
    \\
    $p_{qq} \; \ln^2(1-x)$            & $0$  & $-8$    & $6$     & $-2$    & $0$
   & $0$     & $0$
    \\
    $p_{qq} \;\: \Li(1)$              & $-4$ & $0$     & $-4$    & $-4$    & $0$
   & $0$     & $0$
    \\
    $p_{qq} \;\: \Li(1-x)$            & $0$  & $0$     & $0$     & $0$     & $0$
   & $0$     & $0$
    \\
    $\phantom{(}1-x$                  & $3$  & $22/3$  & $-4$    & $10/3$  &
$-4/3$ & $0$     & $-4/3$
    \\
    $(1-x) \; \ln{x}$                 & $-4$ & $0$     & $0$     & $0$     & $0$
   & $0$     & $0$
    \\
    $(1-x) \; \ln(1-x)$               & $0$  & $-8$    & $4$     & $-4$    & $0$
   & $0$     & $0$
    \\
    $\phantom{(}1+x$                  & $0$  & $0$     & $0$     & $0$     & $0$
   & $0$     & $0$
    \\
    $(1+x) \; \ln{x}$                 & $0$  & $0$     & $0$     & $0$     & $0$
   & $0$     & $0$
    \\ \hline
\multicolumn{8}{|c|}{\em Single Spurious Poles}
    \\ \hline
    $p_{qq} \; I_0$                   & $0$  & $0$     & $-4$    & $-4$    & $0$
   & $0$     & $0$
    \\
    $p_{qq} \; I_0 \ln{x}$            & $-4$ & $0$     & $-4$    & $-4$    & $0$
   & $0$     & $0$
    \\
    $p_{qq} \; I_0 \ln(1-x)$          & $-4$ & $-8$    & $4$     & $-4$    & $0$
   & $0$     & $0$
    \\
    $p_{qq} \; I_1$                   & $4$  & $0$     & $4$     & $4$     & $0$
   & $0$     & $0$
    \\
    $(1-x) \; I_0$                    & $-4$ & $-8$    & $4$     & $-4$    & $0$
   & $0$     & $0$
    \\ \hline
  \end{tabular}
  \caption{Contributions to the inclusive splitting function
$P_{qq}$ from the graphs (e), (f), and (g) in the NPV
prescription.}
  \label{tb:2}
\end{table}

In addition we present also contributions from diagrams of topologies (d-qq) and
(d-gg) to the {\em exclusive splitting functions} in which dependence on the
final-state momenta is kept unintegrated, see Fig.~\ref{fig:1}:
\begin{align} \label{eq:1}
  & W_{R}^{(\text{d-qq})}
  =
  \as^2 \: \frac{\Gamma(1-\eps)}{(4\pi)^\eps} \: \frac{1}{\abs{k^2}} \:
  \biggl\{
    \frac{1}{\eps} (6 - 4\ln{x} - 8 \ln(1-x) - 16 I_0) \left(
\left(\frac{\abs{k^2}}{\mr^2}\right)^\eps - 1 \right) \Pqq \nonumber \\
    & \qquad + \biggl( \pqq \left(-14 + 16 \Li(1) + 4 \ln^2{x} - 4 \Li(1-x) + 8
I_0 \ln{x} + 8 I_0 \ln(1-x) - 24 I_1 \right) \nonumber\\
    & \hspace{60mm} + (1-x) - (1+x) \biggr)
\left(\frac{\abs{k^2}}{\mr^2}\right)^\eps
  \biggr\}
  \text{,}
\end{align}
\begin{align} \label{eq:2}
  & W_{R}^{(\text{d-gg})}
  =
  \as^2 \: \frac{\Gamma(1-\eps)}{(4\pi)^\eps} \: \frac{8}{\abs{k^2}} \:
  \biggl\{
    \frac{1}{\eps} \left(-\frac{11}{3} + 2\ln{x} + 2 \ln(1-x) + 6 I_0 \right)
\left( 1 - \left(\frac{\abs{k^2}}{\mr^2}\right)^\eps \right) \Pgg \nonumber \\
    & \qquad -
    \left( \Pgg \left(\frac{67}{9} - 6 \Li(1) - 2 \ln^2{x} - 4 I_0 \ln{x} - 2
I_0 \ln(1-x) + 8 I_1 \right) + \frac{x}{6} \right)
    \left(\frac{\abs{k^2}}{\mr^2}\right)^\eps
  \biggr\}
  \text{,}
\end{align}
where the leading-order splitting functions are given as
\begin{align}
  \pqq = \frac{1+x^2}{1-x} \text{,} && \Pqq = \frac{1+x^2}{1-x} + \eps (1-x)
\text{,} && \Pgg = \frac{(1-x+x^2)^2}{x(1-x)}
  \text{.}
\end{align}
Note that contributions to the exclusive splitting functions in
eqs.~(\ref{eq:1}--\ref{eq:2})
contain $1/\eps$ poles in $m=4+2\eps$ dimensions, but they are finite in
$\eps\to0$ limit.
This is not true for the same contributions in the standard PV prescription
where finiteness
for $\epsilon\to0$ occurs only after adding the real contributions%
    \footnote{There are exceptions to this behavior (finiteness for
$\epsilon\to0$)
    also in the NPV scheme: they occur for the topologies (f) and (g) and they
are connected
    with building up of the running coupling constant and with the
final-state-type singularities.}.

\section{Summary}

In this work we have presented all contributions to the non-singlet space-like
NLO splitting function $P_{qq}$ calculated in the NPV prescription, which is a
modification of the standard approach proposed by Curci, Furmanski and
Petronzio \cite{CFP80}.
The main reason for such a modification is the need to have the {\em exclusive}
NLO splitting functions.
These objects serve as building blocks for the NLO parton shower Monte Carlo
generator, \texttt{KRKMC}, currently developed by the theory group at IFJ
PAN \cite{JKPSS13}.

The key modification was made to the regularization prescription for spurious
singularities which arise from axial-type denominators \cite{GJKS14}.
The obtained exclusive results \cite{GJKS14,Git14} have simpler pole structure
(compared with \cite{Hei98}),
and apart from known exceptions they satisfy the requirement for being finite in
$\eps\to0$ limit,
and thus are suitable for Monte Carlo parton shower simulations in four
dimensions.
We also presented the inclusive results separately for real and virtual
contributions.
These results are in full agreement with the literature
\cite{CFP80,EV96,Hei98}.

  \section{Acknowledgments}

This work is partly supported by 
 the Polish National Science Center grants DEC-2011/03/B/ST2/02632
and UMO-2012/04/M/ST2/00240,
the U.S.\ Department of Energy
under grant DE-FG02-13ER41996, the Lightner-Sams Foundation,
and the European Commission through contract PITN-GA-2010-264564 (LHCPhenoNet).


\begin{thebibliography}{10}

\bibitem{AP77}
G. Altarelli and G.~Parisi.
\newblock {Asymptotic Freedom in Parton Language}.
\newblock {\em Nucl.Phys.}, B126:298, 1977.

\bibitem{Gri72}
V.N. Gribov and L.N. Lipatov. 
\newblock {Deep inelastic e p scattering in perturbation theory}.
\newblock {\em Sov. J. Nucl. Phys.} {\bf 15} (1972) 438.

\bibitem{Dok77}
Y.L. Dokshitzer.
\newblock {Calculation of the Structure Functions for Deep Inelastic Scattering
  and e+ e- Annihilation by Perturbation Theory in Quantum Chromodynamics.}
\newblock {\em Sov.Phys.JETP}, 46:641--653, 1977.

\bibitem{CFP80}
G.~Curci, W.~Furmanski, and R.~Petronzio.
\newblock {Evolution of parton densities beyond leading order: the nonsinglet
  case}.
\newblock {\em Nucl.Phys.}, B175:27, 1980.

\bibitem{FRS77}
E.G.~Floratos, D.A.~Ross, and C.T.~Sachrajda.
\newblock {Higher order effects in asymptotically free gauge theories:
the anomalous dimensions of Wilson operators}.
\newblock {\em Nucl.Phys.}, B129:66, 1977.

\bibitem{FRS79}
E.G.~Floratos, D.A.~Ross, and C.T.~Sachrajda.
\newblock {Higher order effects in asymptotically free gauge theories.
2. flavor singlet Wilson operators and coefficient functions}.
\newblock {\em Nucl.Phys.}, B152:493, 1979.

\bibitem{FP80}
W.~Furmanski and R.~Petronzio.
\newblock {Singlet Parton Densities Beyond Leading Order}.
\newblock {\em Phys.Lett.}, B97:437, 1980.

\bibitem{EGMPR79}
R.K. Ellis, Howard Georgi, M. Machacek, H.D. Politzer, and G.G.
  Ross.
\newblock {Perturbation Theory and the Parton Model in QCD}.
\newblock {\em Nucl.Phys.}, B152:285, 1979.

\bibitem{MVV04a}
S.~Moch, J.A.M. Vermaseren, and A.~Vogt.
\newblock {The Three loop splitting functions in QCD: The Nonsinglet case}.
\newblock {\em Nucl.Phys.}, B688:101--134, 2004.

\bibitem{MVV04b}
A.~Vogt, S.~Moch, and J.A.M. Vermaseren.
\newblock {The Three-loop splitting functions in QCD: The Singlet case}.
\newblock {\em Nucl.Phys.}, B691:129--181, 2004.

\bibitem{AMV12}
A.A. Almasy, S.~Moch, and A.~Vogt.
\newblock {On the Next-to-Next-to-Leading Order Evolution of Flavour-Singlet
  Fragmentation Functions}.
\newblock {\em Nucl.Phys.}, B854:133--152, 2012.

\bibitem{MMV06}
A.~Mitov, S.~Moch, and A.~Vogt.
\newblock {Next-to-Next-to-Leading Order Evolution of Non-Singlet Fragmentation
  Functions}.
\newblock {\em Phys.Lett.}, B638:61--67, 2006.

\bibitem{GJKPSS11}
M.~Skrzypek, S.~Jadach, A.~Kusina, W.~Placzek, M.~Slawinska, and O.~Gituliar.
\newblock {Fully NLO Parton Shower in QCD}.
\newblock {\em Acta Phys.Polon.}, B42:2433--2443, 2011.

\bibitem{JKPSS13}
S.~Jadach, A.~Kusina, W.~Placzek, M.~Skrzypek, and M.~Slawinska.
\newblock {Inclusion of the QCD next-to-leading order corrections in the
  quark-gluon Monte Carlo shower}.
\newblock {\em Phys.Rev.}, D87:034029, 2013.

\bibitem{JSKS09}
S.~Jadach, M.~Skrzypek, A.~Kusina, and M.~Slawinska.
\newblock {Exclusive Monte Carlo modelling of NLO DGLAP evolution}.
\newblock {\em PoS}, RADCOR2009, 069, 2010.

\bibitem{OMR13}
E.G.~Oliveira, A.D.~Martin, and M.G.~Ryskin.
\newblock {Treatment of the infrared contribution: NLO QED
evolution as a pedagogic example}.
\newblock {\em Eur.Phys.J.}, C73, 2534, 2013.

\bibitem{Oliveira13}
E.G.~Oliveira, A.D.~Martin, and M.G.~Ryskin.
\newblock {Physical factorisation scheme for PDFs for non-inclusive
applications}.
\newblock {\em JHEP}, 1311, 156, 2013.

\bibitem{EV96}
R.K. Ellis and W.~Vogelsang.
\newblock {The evolution of parton distributions beyond leading order: the
  singlet case}.
\newblock 1996.

\bibitem{Hei98}
G. Heinrich.
\newblock {\em {Improved techniques to calculate two-loop anomalous dimensions
  in QCD}}.
\newblock PhD thesis, Swiss Federal Institute of Technology, Zurich, 1998.

\bibitem{JKSS11}
S.~Jadach, A.~Kusina, M.~Skrzypek, and M.~Slawinska.
\newblock {Two real parton contributions to non-singlet kernels for exclusive
  QCD DGLAP evolution}.
\newblock {\em JHEP}, 1108:012, 2011.

\bibitem{GJKS14}
O.~Gituliar, S.~Jadach, A.~Kusina, and M.~Skrzypek.
\newblock {On regularizing the infrared singularities in QCD NLO splitting
  functions with the new Principal Value prescription}.
\newblock {\em Phys.Lett.}, B732:218, 2014.

\bibitem{Git14}
O. Gituliar.
\newblock {\em {Higher-Order Corrections in QCD Evolution Equations and Tools
  for Their Calculation}}.
\newblock PhD thesis, Institute of Nuclear Physics, Polish Academy of Sciences,
  Cracow, 2014.

\bibitem{Axiloop}
O. Gituliar {\it et al.},
{\tt Axiloop} package, http://gituliar.org/axiloop/.

\end{thebibliography}
\end{document}